\documentclass[prd,showpacs,nofootinbib,preprint, 10 pt]{revtex4}
\usepackage[a4paper, total={7in, 10in}]{geometry}
\usepackage{amsmath,graphicx,xcolor,epsfig,slashed}
\usepackage{appendix}
\usepackage{tabularx}
\usepackage{float}
\usepackage{pdfpages}
\usepackage{epstopdf}
\usepackage{subfigure}
\usepackage{caption}
\usepackage{multirow}
\usepackage{booktabs,siunitx}
\usepackage{hyperref}
\usepackage{orcidlink}

\begin{document}
\begin{center}
{\bf\Large\boldmath
Radiative Higgs decay to $\psi\left(4230\right)$}\\[5mm]
\par\end{center}

\begin{center}
\setlength{\baselineskip}{0.2in}
Ahmed Ali\,\orcidlink{0000-0002-1939-1545}, \\
Deutsches Elektronen-Synchrotron DESY, Notkestr. 85, 22607 Hamburg, Germany\\
Muhammad Jamil Aslam \,\orcidlink{0000-0001-9192-066X}, \\
Department of Physics, Quaid-i-Azam University, Islamabad, 45320, Pakistan
\end{center}
\textbf{Abstract}\\[5mm] 

We calculate the decay rate for the  radiative Higgs decay $H \to \psi(4230) + \gamma$ within and beyond the SM. The vector  state $\psi(4230)$
is a candidate exotic tetraquark state and has the quantum
numbers $I^G(J^{PC})=0^-(1^{--})$, having a mass $m_{\psi(4230)}=4221.7 \pm 2.5$ MeV and a decay width $\Gamma_{\psi(4230)}=51 \pm 7$ MeV.
We propose  a practical parametrization of the so-called {\it direct amplitude}, using the leptonic coupling of the
$\psi(4230)$ to estimate the short-distance overlap relevant to the Higgs decay amplitude. We find 
${\cal B}_{\rm SM}( H \to \psi(4230)  \gamma)=4.88^{+4.71}_{-4.52}\times 10^{-7}$, where  the uncertainties are dominated by the experimental uncertainty in the leptonic decay width of the $\psi(4230)$.
The branching ratio increases to about  $10^{-6}$ for $y_c=-3 $, where $y_c$ is the Higgs-Yukawa coupling of the charm quark.


\section{Introduction}\label{sec:1}

\begin{figure}[!htb]
\centering\scalebox{1}{
\includegraphics[width=5.0in,height=2.2in]{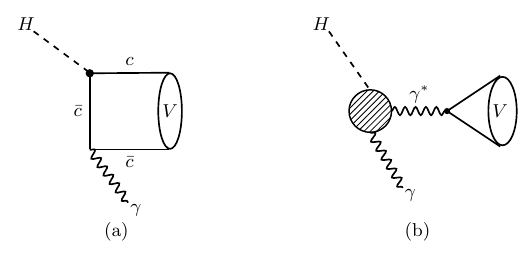}}
\caption{Feynman diagram for (a) direct, and (b) indirect amplitudes for the decay $H\to V\gamma$, where $V=J/\psi$.}
\label{FeynDiag1}
\end{figure}

Radiative decays of the Higgs boson to the vector quarkonium states $H \to J/\psi \gamma$ and $H \to \Upsilon(nS) \gamma$ have been proposed as a measure of the Higgs Yukawa  couplings $HQ\bar{Q}$, $y_Q$, where $Q$ is either a charm quark (for the $J/\psi$) or a bottom quark 
(for the $\Upsilon(nS), n=1,2,3)$~\cite{Keung:1983ac}. In the case of $H \to J/\psi \gamma$,
the Higgs couples to the $c\bar{c}$ pair, and the photon is radiated before the $c\bar{c}$  forms the final state $J/\psi$.  
This is called the {\it direct production} mechanism, and the corresponding amplitude depends on the coupling $y_c$.
 The next-to-leading order (NLO) correction to this process was calculated
in~\cite{Vysotsky:1980cz}, and the leading logarithms were resummed in~\cite{Shifman:1980dk}.
In~\cite{Jia:2008ep,Bodwin:2013gca}, an {\it indirect production} mechanism was introduced, which proceeds through the $H \to \gamma \gamma^*$ decay, with the subsequent transition $\gamma^* \to J/\psi$. The two mechanisms for the decay $H \to J/\psi \gamma$ are shown in Fig.~\ref{FeynDiag1}.
Using NRQCD~\cite{Bodwin:1994jh}, at leading order (LO) in the strong coupling constant $\alpha_s$ and the heavy-quark  velocity $v_c$, the amplitudes for $H\to V\left(=J/\psi, \Upsilon\left(1S\right)\right)\gamma$ were  recalculated in ~\cite{Bodwin:2013gca}. 
 Subsequently, these decays  have been calculated with improved accuracy in a number of theoretical approaches
\cite{Konig:2015qat, Bodwin:2016edd,Brambilla:2019fmu}. One such improved calculation yields
 ${\cal B}_{\rm SM}( H \to J/\psi \gamma)=(3.01\pm 0.15)\times 10^{-6}$ and ${\cal B}_{\rm SM}( H \to  \Upsilon\left(1S\right)\gamma) =(9.97 ^{+4.04}_{-3.03})\times 10^{-9}$~\cite{Brambilla:2019fmu}. The decay $H \to \psi(2S) \gamma$ is estimated to have the branching ratio 
 ${\cal B}_{\rm SM}( H \to \psi(2S) \gamma)=(1.03 \pm 0.06)\times 10^{-6}$~\cite{dEnterria:2023wjq}. Searches for the rare decays of the Higgs and the $Z$, $(H,Z) \to V \gamma$, for $V=J/\psi$ and $\psi(2S)$, with the subsequent decay 
$J/\psi \to \mu^+\mu^-$~\cite{ParticleDataGroup:2026mpi}, have been carried out by the ATLAS and CMS collaborations at the Large Hadron Collider (LHC). No evidence for these decays has been found so far. The results of these searches can be summarised in
terms of the 95\% C.L. upper limits. Based on 139 fb$^{-1}$ integrated luminosity collected at the proton-proton center-of mass energy of 13 TeV, ATLAS has quoted
 ${\cal B}(H \to J/\psi \gamma) < 2.0 \times 10^{-4}$ and ${\cal B}(H \to \psi(2S)  \gamma) < 10.5 \times 10^{-4}$~\cite{ATLAS:2022rej}. The corresponding upper limits from the CMS collaboration are  
${\cal B}(H \to J/\psi \gamma) < 2.6 \times 10^{-4}$ and ${\cal B}(H \to \psi(2S)  \gamma) < 9.9 \times 10^{-4}$~\cite{CMS:2024hhg}. These upper limits are about two (three) orders of magnitude above
SM expected rates for $H \to J/\psi \gamma$ ($H \to \psi(2S)  \gamma$). 
 These decay rates are sensitive to the Yukawa couplings $y_c$.

\vspace*{3.0mm}
  In this Letter, we report on the decay rates for the radiative Higgs decay $H \to \psi(4230) + \gamma$ in the SM and beyond. The vector  state $\psi(4230)$ has the quantum
numbers $I^G(J^{PC})=0^-(1^{--})$, having a mass $m_{\psi(4230)}=4221.7 \pm 2.5$ MeV and a decay width $\Gamma_{\psi(4230)}=51 \pm 7$ MeV\cite{ParticleDataGroup:2026mpi} . Originally called $Y(4260)$, it  was first reported by BaBar in the initial state radiation 
$e^+e^- \to (\gamma_{\rm ISR}) Y(4260)$ \cite{BaBar:2005hhc}. Its properties were measured with much improved precision by BESIII, but
to be well described their data required another nearby state
at 4320 MeV ~\cite{BESIII:2016bnd}. Fitting the BESIII data with two Breit-Wigner distributions led to a narrower structure, called $\psi(4230)$ in the PDG 
Listings~\cite{ParticleDataGroup:2026mpi}. There are presumably yet more charmonium-like exotic vector states
present in the data -  an active area of research at the BESIII and BelleII experiments (for reviews see~\cite{ParticleDataGroup:2026mpi,Ali:2017jda,Olsen:2017bmm}).
%
 From the discovery decay mode $\psi(4230) \to J/\psi \pi^+ \pi^-$ \cite{BaBar:2005hhc},  as well as from 
a number of similar decay modes, such as $\psi(4230) \to \psi(2S) \pi^+ \pi^-,\,h_c(1P) \pi^+\pi^-$~\cite{ParticleDataGroup:2026mpi}, 
the $\psi(4230)$ is in all likelihood a tetraquark, with the quark configuration $c\bar{c} q\bar{q}$, where $q=u,d$ quarks. 
 Its leptonic decay width has been measured, $\Gamma_{ee}[\psi(4230)]= 1.53 \pm 0.26 \pm 0.54$ keV, which is of the same order as $\Gamma_{ee}[\psi(2S)]= 2.29 \pm 0.06$ keV.  Its leptonic branching ratio is small, making the searches in the decay mode $\psi(4230)\to \ell^+ \ell^-$ very challenging.  However, a good number of the $\psi(4230)$ decay modes are already measured at the $e^+e^-$ collider experiments~\cite{ParticleDataGroup:2026mpi}, which  are also amenable 
  at the LHC. Involving charged particle in the final state,  the mass constraints provided by  the Higgs and the $\psi(4230)$ would bring the hadronic
  background under control. Among the candidate exotic vector states in the $c \bar{c}$ sector, $\psi(4230)$ is the best measured so far.  We calculate the
   branching ratio for the decay $H \to \psi(4230) \gamma$ here, expanding the exclusive radiative decays of the Higgs boson to tetraquarks.

\vspace*{3.0mm}
The amplitudes for the decays $H \to \psi(4230) \gamma$ and $H \to J/\psi \gamma$ have some common features, but
they differ in a crucial way from each other. The similar features are in the {\it indirect amplitudes}, which involve the decay $H \to \gamma \gamma^*$ followed by the
coupling $\gamma^* \to V$, where $V$ is either the $J/\psi$ or the $\psi(4230)$. The crucial difference  lies in the {\it direct amplitudes}. The Higgs-Yukawa interaction ${\cal L}_Y =-m_c/v H\bar{c}c$ creates a short-distance charm-anticharm quark pair. For a vector charmonium, such as $J/\psi$, this pair can be projected onto a conventional quarkonium wave function, as shown in Fig.~\ref{FeynDiag1}. Alternatively, one can use the QCD factorization approach~\cite{Beneke:1999br}, based on heavy to light transitions,
using the light-cone distribution amplitudes (LCDA) and perturbative kernels. 
 For an exotic tetraquark state, such as $\psi(4230)$, neither the corresponding LCDAs nor the perturbative kernel are  known. More importantly,  
 the relevant factor needed is not simply the probability of finding charm in the state. Hidden charm-molecules~\cite{Wang:2013cya} and compact tetraquarks~\cite{Maiani:2004vq,Maiani:2014aja,Ali:2017wsf}, the two competing theoretical frameworks for
the  $\psi(4230)$, necessarily contain the $c$ and $\bar{c}$, but the Higgs amplitude probes their ability to match onto a locally produced short-distance $c\bar{c}$ configuration. In the approximation we are working (lowest order in the QCD coupling $\alpha_s(\mu)$), the {\it direct amplitude}. can be written in the factorized form:
\begin{equation}
{\cal A}_{\rm dir} (H \to \psi(4230) \gamma) = A(H \to \gamma c\bar{c}) {\cal M}(c\bar{c} \to \psi(4230))\;. \label{eq:2}
\end{equation}
The Higgs amplitude $A(H \to \gamma c\bar{c})$ is well-known, but the quantity ${\cal M}(c\bar{c} \to \psi(4230))$ is an {\it a priori} unknown short-distance overlap. This distinction (between a compact $J/\psi$ and a $\psi(4230)$) is specially important for a spatially extended molecule. For the case of $\psi(4230)$ as a compact tetraquark, the production is a fragmentation process.  Once the $c\bar{c}$ is produced by the Higgs, it picks up a light $q\bar{q}$ pair from the vacuum to make the vector state with the quark configuration $c \bar{c} q \bar{q}$. One could attempt
to calculate ${\cal M}(c\bar{c} \to \psi(4230))$ in a detailed model for the underlying dynamics. This should be undertaken in future theoretical works. However, here we propose a practical parametrization for the $H \to \psi(4230) \gamma$ decay using the leptonic coupling as a data-driven short-distance probe. 

\vspace*{3.0mm}
The details of our calculation are given below. The
current measurements of the $J/\psi$ and the $\psi(4230)$ states, together with other input parameters, are  given in Table~\ref{tab:MesonPar}.
Representative diagrams for the $\psi(4230)$  as a $\bar{D}_1 D$ molecule are shown in Fig.~\ref{FeynDiag2} and as a compact tetraquark $[cq] [\bar{c}\bar{q}]$ 
are shown in Fig.~\ref{FeynDiag3}, respectively. 
 The contributions from the direct and indirect amplitudes, as well as the decay rates,  are given in Table \ref{tab:rate_BR}. We  find ${\cal B}_{\rm SM}( H \to \psi(4230)  \gamma)=4.88^{+4.71}_{-4.52}\times 10^{-7}$, with the errors dominated by the leptonic decay width of the $\psi(4230)$.
 In some beyond-the-SM scenarios, the Higgs-Yukawa charm quark
coupling may deviate from its SM value. We show the deviation  of ${\cal B}\left( H \to \psi(4230)\gamma\right)$ for the range $-3 \leq y_c \leq 3$
in Fig.\ref{FeynDiag4} ($y_c=1$ in the SM) which lead to a factor of $2$ increase for $y_c = -3$.  

\section{Calculation Details}\label{sec:2}

\begin{figure}[!htb]
\centering\scalebox{1}{
\begin{tabular}{cc}
\includegraphics[width=3.0in,height=2.0in]{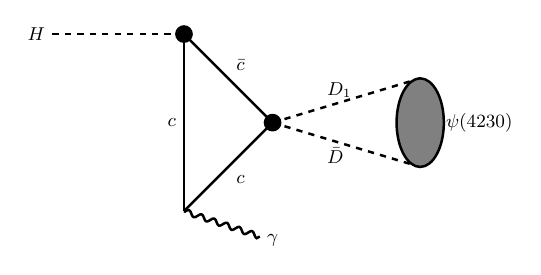}\hspace{0.5cm}\qquad & \qquad
\includegraphics[width=3.0in,height=2.0in]{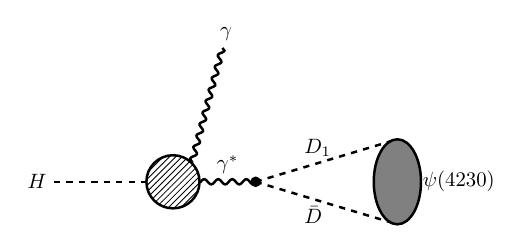} \\
(a)\hspace{0.5cm}\qquad &  \qquad (b)
\end{tabular}}
\caption{Feynman diagram for (a) $H \to c\bar{c} \gamma \to D_1 \bar{D} \gamma \to \psi(4230) \gamma$, and (b) $ H \to \gamma \gamma^* \to
D_1 \bar{D} \gamma \to \psi(4230) \gamma$.}
\label{FeynDiag2}
\end{figure}

There are two competing theoretical frameworks for the tetraquark $\psi(4230)$.
 The first, a molecular interpretation, is based essentially on the premise that $D_1(2420) \bar{D}$ is the dominant component  
of the $\psi(4230)$~\cite{Ding:2008gr,Wang:2013cya,Li:2013yla,Cleven:2013mka}. 
The Fock state $\psi(4230)$ in this picture can be described as:
\begin{equation}
\vert \psi(4230) \rangle = \sqrt{Z} \vert c\bar{c}\rangle + \sqrt{1-Z} \vert D_1 \bar{D}
+ \bar{D}_1 D\rangle\;.
\end{equation}
 Here, $Z$ is a measure of the compact $c\bar{c}$ component~\cite{Weinberg:1965zz}, proportional to $\sqrt{Z}$, and the non-compact component 
($ D_1 \bar{D}+ \bar{D}_1 D$) is proportional to $\sqrt{1-Z}$. An analysis of the dipion mass spectrum in
the process $e^+e^- \to \psi(4230) \to J/\psi \pi^+ \pi^-$ finds  that the singlet-component of the $\psi(4230)$ (\textit{i.e.}, $c\bar{c}$) is small, and $\psi(4230)$ has a large flavor-octet light-quark component (\textit{i.e.}, $ D_1 \bar{D} + \bar{D}_1 D $)~\cite{ Chen:2019mgp}. There may also be other subdominant contributions
present in the $\psi(4230)$~\cite{ Chen:2019mgp}. Representative diagrams depicting the direct and indirect decay
$H \to \psi(4230) \gamma$ in the molecular interpretation of the $\psi(4230)$ are shown in Fig.~\ref{FeynDiag2}.

The compact tetraquark picture in the $c\bar{c}$ sector has four orbitally excited vector states, called $Y$ tetraquarks. They are all $P (L=1)$ states, having $J^{PC}=1^{--}$,
built  from the diquark ${\it Q}=[cq]_{\bar{3}}$ and anti-diquarks ${\bar{\it Q}}=[\bar{c}\bar{q}]_3$, where the subscript denotes color representation. These states
are characterized by the quantum numbers $\vert S_{\it Q}, S_{\bar{\it Q}}, { S}, { L}\rangle_J$, where $ S_{\it Q}, (S_{\bar{\it Q}}) $ are the spins of the diquark(antidiquark),
with ${\bf S}= { \bf S}_Q+ {\bf S}_{\bar{Q}} $, $ L$ is the orbital quantum number with $\bf J= L + S$. It has been argued in~\cite{Ali:2017wsf} that $\psi(4230)$
is the lowest-lying orbitally excited state with the  quantum numbers $\vert 0, 0, 0, 1\rangle_1$. The color-antitriplet scalar diquark can be written as
\begin{equation}
S^{a}(x)= \epsilon^{abc} q_b^T(x) C \gamma_5 c_c(x)\; ,\label{eq:S1}
\end{equation}
where  $\epsilon^{abc}$ is the totally antisymmetric Levi-Civita tensor in the quark color space, $C$ is the charge-conjugation matrix,
the superscript $T$ denotes transposition, and  the color indices are summed. The corresponding color-triplet antidiquark is
\begin{equation}
\bar{S}^{a}(x)= \epsilon^{abc} \bar{q}_b (x) \gamma_5 C \bar{c}^{T}_c(x)\; .\label{eq:S2}
\end{equation}

\begin{figure}[!htb]
\centering\scalebox{1}{
\includegraphics[width=5.0in,height=2.2in]{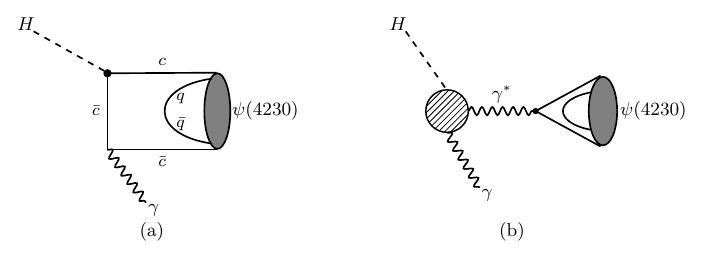}}
\caption{Feynman diagram for (a) direct, and (b) indirect amplitudes for the decay $H\to \psi\left(4230\right)\gamma$  in the
compact tetraquark picture. The light $q\bar{q}$ pair is produced non-perturbatively.}
\label{FeynDiag3}
\end{figure}

 A scalar-scalar $J^{PC}= 1^{--}$ current in this picture has the compact representation
\begin{equation}
 J_\mu^{\rm (SS)} = S^{a} (x) \overleftrightarrow{D}_\mu \bar{S}^{a} (x)  \;, \label{eq:Jdef1}
\end{equation}
where the covariant derivative is $S\overleftrightarrow{D}_\mu \bar{S} = S ( D_\mu \bar{S}) - (D_\mu S) \bar{S}$. Equivalently, using the quark fields,
\begin{equation}
 J_\mu^{\rm (SS)} = \epsilon^{abc}\epsilon^{ade} [q_b^T C \gamma_5 c_c] \overleftrightarrow{D}_\mu [\bar{q}_d \gamma_5 C \bar{c}^{T}_e   ] \;. \label{eq:Jdef2}
\end{equation}
This is a gauge-invariant local dimension-7 current. 
Schematic diagrams depicting the direct and indirect decay
$H \to \psi(4230) \gamma$ in the compact tetraquark interpretation of the $\psi(4230)$ are shown in Fig.~\ref{FeynDiag3}. There is no direct coupling of the Higgs, or the photon, with the composite objects ${\it Q}$  and ${\bar{\it Q}}$. As discussed below, we are interested in the electromagnetic current matrix element
 $\langle 0 \vert J_\mu \vert \psi(4230\rangle$.


%

\vspace*{3.0mm}
\subsection{Direct Amplitude}
Formally, the direct amplitude for the decay $\mathcal{A}_{\mathrm{dir}}\left(H\to \psi(4230)\gamma\right)$  can be expressed like the corresponding well-known leading-order NRQCD amplitude, derived for conventional charmonium 
$\mathcal{A}_{\mathrm{dir}}\left(H\to J/\psi\gamma\right) $~\cite{Bodwin:2013gca} as: 
\begin{equation}
\mathcal{A}_{\mathrm{dir}}\left(H\to \psi\left(4230\right)\gamma\right)= 2\sqrt{3}e_c ey_{c}\left(2\sqrt{2}G_{F}\frac{m_{c}}{m_H}\right)^{1/2}\frac{1-\frac{m_{\psi(4230)}^{2}}{m_{H}^{2}}}{\left(1-\frac{m_{\psi(4230)}^{2}}{2m_{H}^{2}}-\frac{2m_{c}^{2}}{m_{H}^{2}}\right)} \phi_0\left[{\psi(4230)}\right]\;.\label{eq:4a}
\end{equation}
 In Eq. (\ref{eq:4a}), $e_c$ is the electric charge of the charm quark, $e$ is the electromagnetic coupling, and $G_F$ is the Fermi coupling constant. 
  The Higgs-Yukawa charm
 coupling $(g_{Hc\bar{c}}$) is allowed to differ from its SM value. We introduce the dimensionless Yukawa coupling modifier, $y_c$, defined by
  $g_{Hc\bar{c}}= y_c (g_{Hc\bar{c}})_{\rm SM}$, so that $y_c=1$ in the SM. Both positive and negative values of $y_c$ are considered.
   For the decay $H \to \psi(4230)\gamma$,  the quantity $\phi_0\left[\psi(4230)\right]$ is the
 overlap discussed in the preceding section. To evaluate it, we note that a closely related phenomenological quantity is the coupling of the $\psi(4230)$ to a local charm current
 \begin{equation}
 \langle 0 \vert \bar{c} \gamma_\mu c \vert \psi(4230)\rangle= g_{\psi(4230)\gamma}  \epsilon_\mu, \label{eq:fpsi}
 \end{equation}
 where $\epsilon_\mu$ is the polarization vector for the $\psi(4230)$. For an exotic state $\psi(4230)$, it measures, in a directly useful way, the ability of a
 local $c\bar{c}$ current to create the physical $\psi(4230)$ state. This is much closer, conceptually, to the overlap required by a short-distance Higgs process than the
 total charm content of the wave function.
 
 \vspace*{3.0mm}
 We note that Eq.~(\ref{eq:fpsi}) defines a matrix element of the charm-quark vector current, whereas the measured electronic width constrains the full
 electromagnetic current, defined as $J_\mu= \sum_q Q_q \bar{q}\gamma_\mu q$. For a conventional $c \bar{c}$ state, charm-current dominance is a 
 natural assumption. For a tetraquark with light-quark constituents, the electromagnetic current can also couple through the light-quark component.
 However, for the decay $H \to \psi(4230) \gamma$, the contribution to the direct amplitude,  in which the Higgs couples to the light $q\bar{q}$ pair and the $c\bar{c}$ is produced from the vacuum, is comparatively negligible. While 
 extracting the charm-current matrix element directly from $\Gamma_{ee}$ requires an additional assumption, namely the dominance of the charm vector
 current, it is expected to hold for the direct amplitude.
   
 \vspace*{3.0mm}
 With this assumption,
 the electronic width $ \Gamma_{ee}\left(\psi (4230)\right) $ is related to the coupling $ g_{\psi(4230)\gamma}$ by
\begin{equation}
\Gamma_{ee}\left(\psi (4230)\right)= \frac{4 \pi \alpha^2 g_{\psi(4230)\gamma}^2}{3 m_{{\psi(4230)}}^3}\;, \label{eq:5a}
\end{equation}
where $\alpha$ is the fine structure constant and its value at $m_{\psi(4230)}^2$ is given in Table \ref{tab:MesonPar}. 
This prescription provides an effective normalization of the short-distance $c\bar c$ contribution to the direct amplitude.

\vspace*{3.0mm}
The overlap $\phi_0\left[\psi(4230)\right]$, appearing in Eq.~(\ref{eq:4a}) is related to $ g_{\psi(4230)\gamma}$
by \cite{Bodwin:2013gca}
\begin{equation}
\phi_0[\psi(4230)]  =- \frac{1} {2e_c\sqrt{N_c}} \frac{g_{\psi(4230)\gamma}}
{\sqrt{ m_{\psi(4230)}}}\;, \label{eq:6}
\end{equation}
where $N_c=3$, and we have kept the same convention (and sign) as one encounters in the NRQCD approach to the $H \to J/\psi \gamma$ case. For comparison, the 
numerical value of $\phi^2_0\left(J/\psi\right) $  for the decay $H \to J/\psi \gamma$ is given in Table~\ref{tab:MesonPar}.


\vspace*{3.0mm}
For the exotic state $\psi(4230)$, the relation in Eq. (\ref{eq:6}) is adopted as a phenomenological prescription, motivated by the corresponding NRQCD relation for conventional charmonium. The leptonic coupling $(g_{\psi(4230)\gamma})$ provides an experimentally constrained normalization of the matrix element of the local charm-quark vector current. However, its identification with the nonperturbative overlap entering the direct Higgs decay amplitude is not guaranteed by a factorization theorem for exotic multiquark states. In particular, additional nonperturbative contributions and a possible relative phase between the direct and indirect amplitudes cannot be excluded. Our numerical predictions should therefore be understood as estimates within this phenomenological prescription, rather than model-independent predictions.

\vspace*{3.0mm}
In terms of $g_{\psi(4230) \gamma}$, using Eqs. (\ref{eq:4a}) and (\ref{eq:6}), the direct amplitude takes the form
\begin{equation}
\mathcal{A}_{\mathrm{dir}}\left(H\to \psi\left(4230\right)\gamma\right)= - e\,y_{c}\left(2\sqrt{2}G_{F}\frac{m_{c}}{m_H}\right)^{1/2}\frac{1-\frac{m_{\psi(4230)}^{2}}{m_{H}^{2}}}{\left(1-\frac{m_{\psi(4230)}^{2}}{2m_{H}^{2}}-\frac{2m_{c}^{2}}{m_{H}^{2}}\right)} \frac{g_{\psi(4230)\gamma}}
{\sqrt{m_{\psi(4230)}}}\;.\label{eq:5}
\end{equation}
  
To determine $g_{\psi(4230)\gamma}$, we use  the leptonic decay branching ratio of $\psi\left(4230\right)$, which  is measured with the $\mu^+\mu^-$ final state ~\cite{ParticleDataGroup:2026mpi} 
\begin{equation}
\mathcal{B}\left(\psi\left(4230\right)\to \mu^+\mu^-\right)=\left(3.0\pm 2.7\right)\times 10^{-5}\;.\label{eq:7}
\end{equation}
Assuming lepton-flavor universality, $\Gamma_{ee}[\psi(4230)]$ is approximated by $\Gamma_{\mu\mu}[\psi(4230)]$ and its current value is given in Table \ref{tab:MesonPar} together with other
input parameters used in the numerical calculations. 
We have calculated the value of $g_{\psi(4230)\gamma}$ from the 
electronic
decay widths $\Gamma_{ee}[\psi(4230)]$, 
\begin{equation}
g_{\psi(4230)\gamma} = 0.692^{+0.27}_{-0.50}~{\rm GeV}^2\;.
\end{equation}
The large error in  $ g_{\psi(4230)\gamma}  $ is attributed to the poorly known leptonic width  $\Gamma_{ee}[\psi\left(4230\right)]$, which we expect will be greatly improved
in the future by  BESIII and Belle II, yielding a precise value of this coupling.

\begin{table}[tbp]
\renewcommand{\arraystretch}{1.7}
\begin{tabular}{|c|}
\hline\hline
$ m_{\psi (4230)} = 4221.7 \pm 2.5 ~\mathrm{MeV}$  ,\quad
 $\Gamma_{\psi (4230)} = 51 \pm 7 ~\mathrm{MeV}$ ,\quad
 $ \Gamma_{ee}[\psi (4230)] = (1.53 \pm 1.26 \pm 0.54) ~\mathrm{keV}$ \cite{Ablikim:2020jrn}\\
$\alpha\left(m_{\psi\left(4230\right)}\right)=1/132.05$ , \quad 
$ m_{J/\psi} = 3096.900 \pm 0.006 ~\mathrm{MeV}$  ,\quad
 $\Gamma_{J/\psi} = 92.6 \pm 1.7 ~\mathrm{keV}$ \\
 $ \Gamma_{ee}(J/\psi) = (5.53 \pm 0.11) ~\mathrm{keV}$ ,\quad
$\alpha\left(m_{J/\psi}\right)=1/132.64$ \cite{Bodwin:2013gca} , \quad $m_H = 125.20\pm 0.11~\mathrm{GeV}$, \quad $m_c =  1.4~\mathrm{GeV}$ \cite{Brambilla:2019fmu}
\\
$\Gamma_{H}\left(\gamma\gamma\right) = \left(9.31\pm 0.27\right)\times 10^{-3} ~ \mathrm{MeV}$ , \quad $g_{\psi(4230)\gamma} = 0.692^{+0.27}_{-0.50}~\mathrm{GeV}^2$, \quad $\phi^2_0\left(J/\psi\right)= 0.073^{+0.001}_{-0.009}~ \mathrm{GeV}^3$ 
 \cite{Bodwin:2013gca}\\
\hline\hline
\end{tabular}%
\caption{Numerical values of the input parameters. The masses of the particles and the input parameters for $J/\psi$ are taken from \cite{ParticleDataGroup:2026mpi}.}
\label{tab:MesonPar}
\end{table}

 \section{Indirect Amplitude}
 
 The indirect production proceeds through a virtual photon as shown in Fig. \ref{FeynDiag1}b, where the virtual photon couples to a vector 
 state $V = \psi(4230)$ through the matrix element of the electromagnetic current, defined above in Eq.(\ref{eq:fpsi}).
The decay amplitude for the indirect contribution in $H \to \psi(4230) \gamma$ can be parameterized as
\begin{equation}
   \mathcal{M}_{\mathrm{ind}} =-i g_{\psi\gamma}
\mathcal{M}^{\mu\nu}(H\to\gamma\gamma^*)
\frac{g_{\nu\rho}}{q^2}
\epsilon_\mu^*(k)\langle \psi(4230)\left(\epsilon\right)|J^\rho_{\mathrm{em}}\left(0\right)|0\rangle\;,\label{10}
\end{equation}
 where $k$ and $q=p_{\psi(4230)}$ are the momenta of the real and virtual photons, respectively. Using Eq.~(\ref{eq:fpsi}) above, the indirect production amplitude is given by \cite{Bodwin:2013gca}
 \begin{equation}
\mathcal{M}_{\mathrm{ind}}=-e\frac{g_{\psi(4230)\gamma}}{m_{\psi(4230)}^2}\mathcal{M}\left(H\to \gamma\gamma^*\right)\;,\label{eq:11}
 \end{equation}
 where in the leading order in $m_{\psi(4230)}/m_H$,  $\mathcal{M}\left(H\to \gamma\gamma^*\right) = \mathcal{M}\left(H\to \gamma\gamma\right)$.
 In terms of the decay width $\Gamma\left(H\to \gamma\gamma\right)$ the indirect contribution to the amplitude is:
 \begin{equation}
     \mathcal{A}_{\mathrm{ind}}=\frac{eg_{\psi(4230)\gamma}}{m_{\psi\left(4230\right)}^2}\left(1-\frac{m_{\psi\left(4230\right)}^2}{m_H^2}\right)\left[16\pi\Gamma\left(H\to \gamma\gamma\right)\right]^{1/2} \left[1-\left(\frac{m_{\psi\left(4230\right)}}{183.43\;\mathrm{GeV}}\right)^2\right]\;.\label{eq:12}
 \end{equation}
The last factor in Eq.~(\ref{eq:12})  accounts for the leading correction due to the virtuality ($q^2=m_{\psi(4230)}^2$) of the photon in the loop-induced 
$H\to\gamma\gamma^*$ amplitude. For small $q^2$, the loop-induced form factor $F_{\gamma \gamma^*}(q^2)$ can be expressed as
\begin{equation}
\frac{F_{\gamma \gamma^*}(q^2)} {F_{\gamma \gamma^*}(0)} = 1 + C_\gamma q^2 + O(q^4)\; .\label{eq:ff}
\end{equation}

We follow the numerical parametrization of Bodwin et al.~\cite{Bodwin:2013gca}, who obtained
$C_\gamma=1/(183.43\; \mathrm{GeV})^2$  for the $H \to J/\psi \gamma$ decay, using the SM loop form factors discussed in \cite{Djouadi:1996yq}. 
   For $\psi(4230)$, the correction to the amplitude is approximately $C_\gamma(m_{\psi(4230)}^2)=5.3\times10^{-4}$, which is almost negligible.

The corresponding decay width for the $H\to \psi\left(4230\right)\gamma$ is calculated from the following expression:
\begin{equation}
    \Gamma\left(H\to \psi\left(4230\right)\gamma\right)=\frac{1}{8\pi}\left(1-\frac{m_{\psi(4230)}^2}{m_H^2}\right)\left|\mathcal{A_{\rm dir} + A_{\rm ind}}\right|^2\;.\label{eq:13}
\end{equation}

\vspace*{3mm}
\begin{table}[!tbp]
\centering
\renewcommand{\arraystretch}{2.0}
\begin{tabular}{c|cccc}
\hline\hline
 & Direct \quad\quad & Indirect\quad\quad & Interference \quad
 \quad& Total \\
\hline
Decay Rate (MeV)  & $5.82^{+5.41}_{-5.37}\times 10^{-8} $ \quad\quad & $2.74^{+2.70}_{-2.54}\times 10^{-6}$ \quad\quad & $-8.0^{-7.64}_{+7.38}\times 10^{-7}$ \quad\quad & $2.0^{+1.99}_{-1.85}\times 10^{-6}$ \\
Branching ratio & $1.42^{+1.28}_{-1.31}\times 10^{-8}$ \quad\quad& $6.69^{+6.39}_{-6.18}\times 10^{-7}$ \quad\quad & $-1.95^{-1.81}_{+1.80}\times 10^{-7}$ \quad\quad & $4.88^{+4.71}_{-4.52}\times 10^{-7}$ \\
\hline\hline
\end{tabular}
\caption{Decay rates and branching ratios in the SM for the direct, indirect, interference, and the total contributions for $H\to\psi(4230)\gamma$.}
\label{tab:rate_BR}
\end{table}

Using Eqs.~(\ref{eq:5}) and (\ref{eq:12}) in Eq.~(\ref{eq:13}), together with the input parameters listed in Table~\ref{tab:MesonPar}, we obtain the decay rates and branching ratios for the direct, indirect, interference, and the total contributions within the SM, as summarized in Table~\ref{tab:rate_BR}. The central value of the 
decay rate, $\Gamma_{\rm ind}(H\to\psi(4230)\gamma)=0.274\times10^{-8}\;\mathrm{GeV}$, is about a factor 45 larger than the corresponding direct contribution, while being approximately one-fifth of the indirect rate for $H\to J/\psi\gamma$, $\Gamma_{\rm ind}(H\to J/\psi\gamma)=1.37\times10^{-8}\;\mathrm{GeV}$. The latter suppression can be understood quantitatively from the smaller electromagnetic coupling of $\psi(4230)$. Using the electronic decay width of $J/\psi$, we obtain the magnitude of $g_{J/\psi\gamma}=0.83~\mathrm{GeV}^2$. Comparing this with the central value $g_{\psi(4230)\gamma}=0.692~\mathrm{GeV}^2$ from Table~\ref{tab:MesonPar}, we find
\begin{equation}
\left(\frac{g_{\psi(4230)\gamma}}{g_{J/\psi\gamma}}\right)^2
\simeq 0.695\;,
\end{equation}
corresponding to a suppression by a factor of approximately $1.44$ in the coupling-dependent part of the rate. The indirect decay rate ratio is
\begin{equation}
\frac{\Gamma_{\rm ind}(H\to\psi(4230)\gamma)}
{\Gamma_{\rm ind}(H\to J/\psi\gamma)}
\simeq \frac{0.274}{1.37}\simeq 0.20\;,\label{ratio-rates}
\end{equation}
which is smaller than the coupling-squared ratio because of the different phase-space and kinematic factors associated with the two vector-meson masses. In particular, the mass factor, $R_m=(m_{J/\psi}/m_{\psi(4230)})^4$, contributes a suppression factor of approximately $0.290$. Combining this with the coupling ratio yields the overall suppression by a factor of $0.2$, as shown in Eq. (\ref{ratio-rates}).  
  Consequently, the indirect contribution to $H\to\psi(4230)\gamma$ is suppressed relative to that for $H\to J/\psi\gamma$, and the total SM decay rate is likewise smaller, as shown in 
  Table~\ref{tab:rate_BR}. The decay width and branching ratio for $H \to \psi(4230) \gamma$ depend on the charm quark mass, a feature it shares with the $H \to J/\psi \gamma$ decay,
  We use the charm-quark mass, estimated in~\cite{Brambilla:2019fmu}, calculated from the binding energy of the $J/\psi$ in potential models, yielding $m_c=1.4$ GeV.  With this, we get
\begin{equation}
{\cal B}_{\rm SM}( H \to \psi(4230)  \gamma)=(4.88^{+4.71}_{-4.52})\times 10^{-7}\;. \label{eq:smbr1}
\end{equation}
  The relatively large uncertainties in the  branching ratio arise predominantly from the sizeable experimental uncertainty in the electronic decay width of $\psi(4230)$, which directly affects the determination of $|g_{\psi(4230)\gamma}|$ and, consequently, the indirect amplitude. 
  
  \vspace*{3.0mm}
  We expect our estimates of the decay width
  $\Gamma(H \to \psi(4230) \gamma$ to hold, being dominated by the indirect amplitude, which is relatively well-constrained by the measured electromagnetic coupling 
  together with the experimentally well-constrained $H \to \gamma \gamma^*$ coupling. The direct amplitude, $\mathcal{A}_{\mathrm{dir}}\left(H\to \psi\left(4230\right)\gamma\right)$  in  Eq.~(\ref{eq:5}),  calculated under a phenomenological Ansatz, may receive a multiplicative normalization factor $\rho e^{i\delta}$, modifying the magnitude $\rho$ and the relative phase $\delta$ w.r.t. the indirect
  amplitude  $\mathcal{A}_{\mathrm{ind}}\left(H\to \psi\left(4230\right)\gamma\right)$, giving 
  \begin{equation}
    \Gamma\left(H\to \psi\left(4230\right)\gamma\right)=\frac{1}{8\pi}\left(1-\frac{m_{\psi(4230)}^2}{m_H^2}\right)\left|\rho e^{i\delta} \mathcal{A_{\rm dir} + A_{\rm ind}}\right|^2\;.\label{eq:13b}
\end{equation}
  Using the central contributions in Table \ref{tab:rate_BR}, the branching ratio
for $H \to \psi(4230) \gamma$ can be expressed approximately as
\begin{equation}
\mathcal{B} (H\to \psi\left(4230\right)\gamma)
=
\left[6.69+0.142\rho^2-1.95\rho\cos\delta\right]
\times 10^{-7},
\end{equation}
where the last term assumes our adopted interference sign convention. To illustrate it further, the values of the branching ratio obtained by changing $\rho$ and $\delta$ are given in Table \ref{tab:sensitivity}.
\begin{table}[htbp]
\centering
\renewcommand{\arraystretch}{1.0}
\begin{tabular}{lc}
\hline\hline
Assumption & \hspace{1.2 cm} Branching ratio $\left(10^{-7}\right)$ \\
\hline
$\rho=1,\ \delta=0$ (Our Ansatz)
& \hspace{1.2 cm} $4.88$ \\

$\rho=0$ (indirect only)
& \hspace{1.2 cm} $6.69$ \\

$\rho=1,\ \delta=\pi$
& \hspace{1.2 cm} $8.78$ \\
\hline\hline
\end{tabular}
\caption{Sensitivity of the branching ratio $\mathcal{B} (H\to \psi\left(4230\right)\gamma)$ to the relative magnitude
$\rho$ and the phase $\delta$ of the direct amplitude.}
\label{tab:sensitivity}
\end{table}
  Keeping only the indirect amplitude yields $6.69 \times 10^{-7}$ for
the central value of the branching ratio $\mathcal{B} (H\to \psi\left(4230\right)\gamma)$. 
  Studying the dependence of the decay width on $\rho$ and $\delta$ would quantify the robustness of our prediction. Our Ansatz corresponds to assuming $\rho=1$ and $\delta=0$. 
\begin{figure}[!htb]
\centering\scalebox{1}{
\includegraphics[width=5.0in,height=3.5in]{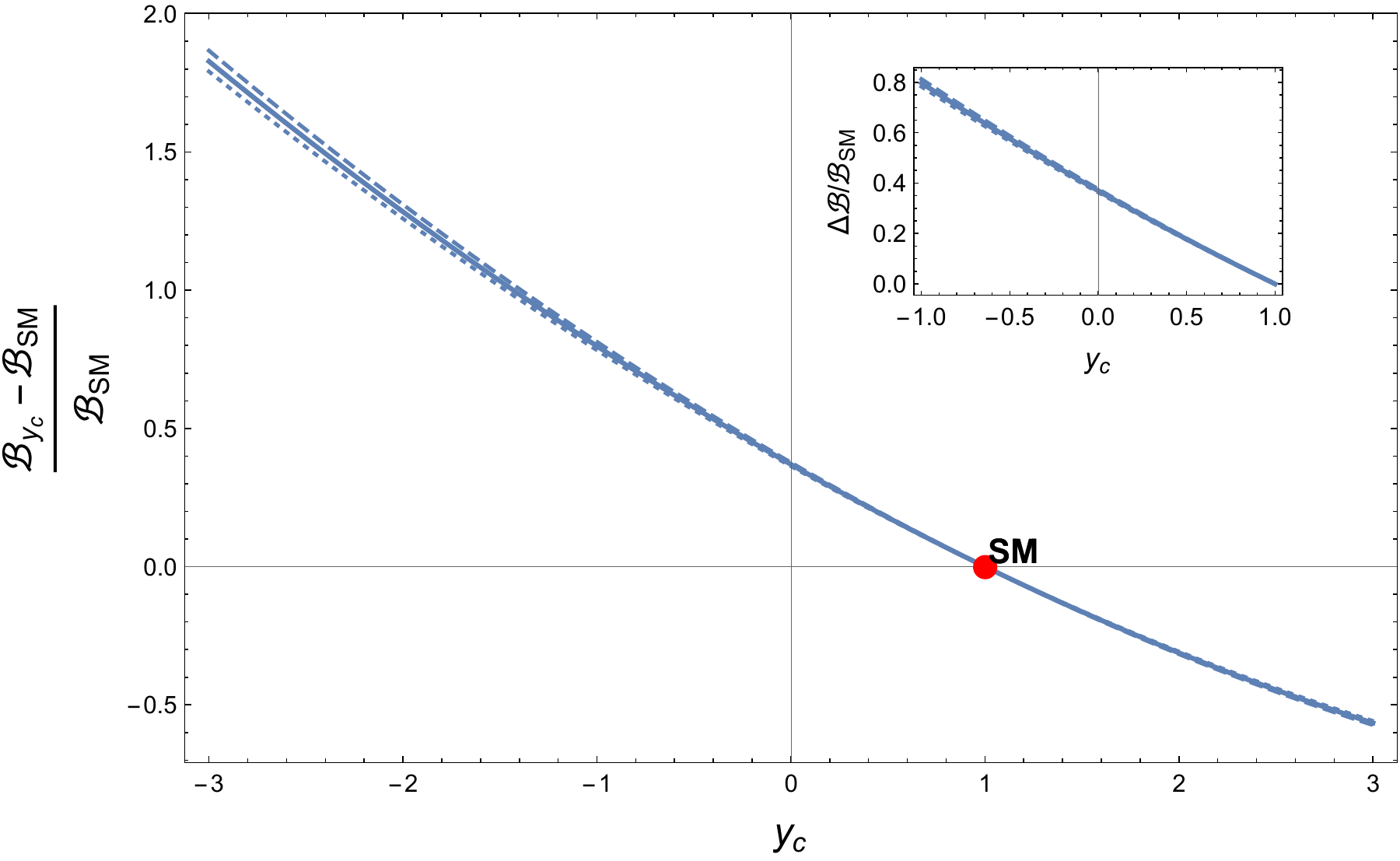}}
\caption{Relative deviations in the branching ratio for $H\to \psi\left(4230\right)\gamma$  as a function of  the Yukawa coupling $y_c$. The solid, dotted and dashed lines correspond to the central, upper and lower value of the input parameters, respectively. 
The solid dot labeled by SM  corresponds to the Standard Model $y_c=1$. }
\label{FeynDiag4}
\end{figure}

\vspace*{3.0mm}
In several beyond-the-standard model scenarios,  an enhanced $y_c$ can be obtained~\cite{Delaunay:2013pja}.
Also, the sign of $y_c$ is not determined.  With that in view, the Higgs Yukawa coupling $y_c$ can be treated as a free parameter and the stated upper limits on ${\cal B}(H \to J/\psi \gamma)$ can be translated
into a bound on $y_c$. 
In general, also the $H\gamma\gamma$ coupling can deviate from its SM value. Parametrizing it as $\kappa_\gamma$, the current measurements are $\kappa_\gamma= 0.97 \pm 0.06$ (ATLAS)\cite{ATLAS:2025qxq}, and $\kappa_\gamma=1.10\pm 0.07$ (CMS)~\cite{CMS:2026zxu}.
These measurements are in agreement with the SM value $\kappa_\gamma=1$.
Using the parametrization~\cite{Brambilla:2019fmu,Perez:2015aoa} 
\begin{equation}
\mu (H \to J/\psi \gamma) = \frac{\vert {\cal A}_{\rm ind} + y_c {\cal A}_{\rm dir}\vert^2}{ \Gamma_{\rm SM} (H \to J/\psi \gamma)   },
\end{equation}
the observed interval for $y_c$ at 95\% C.L. reported by the ATLAS collaboration is $(-133, +175)$ \cite{ATLAS:2022rej}, and the corresponding interval by the CMS collaboration is
 $(-166, +208)$~\cite{CMS:2024hhg}.

 To work out the $y_c$-dependence of  ${\cal B}(H\to\psi(4230)\gamma)$, we write the decay amplitude, following the   $H \to J/\psi \gamma$ case discussed above,  as
\begin{equation}
\mathcal{A}(H\to\psi(4230)\gamma)
=
 \mathcal{A}_{\rm ind} (H\to\psi(4230)\gamma)+   y_c\,\mathcal{A}^{\mathrm{SM}}_{\rm dir}(H\to\psi(4230)\gamma)\;,
\end{equation}
where $\mathcal{A}^{\mathrm{SM}}_{\rm dir} (H\to\psi(4230)\gamma)$ is given in Eq. (\ref{eq:5}) with $y_c=1$. 

The decay rate is now sensitive to both the magnitude and the sign of $y_c$:
\begin{equation}
\Gamma\propto
\left| \mathcal{A}_{\rm ind} +
y_c\,\mathcal{A}^{\mathrm{SM}}_{\rm dir}
\right|^2.
\end{equation}

Consequently, even a moderate departure of $y_c$ from its SM value can modify the branching ratio through the interference term.  The relative branching ratio as a function of $y_c$ is plotted in Fig. ~\ref{FeynDiag4}, where the branching ratio increases by about a factor $2$,  for $y_c=-3 $. 

\vspace*{3mm}
\section{Summary}
In summary, we have investigated the radiative Higgs decay $H\to\psi(4230)\gamma$ within and beyond the SM.
 The vector state $\psi(4230)$ is a candidate exotic tetraquark with the quark configuration $c\bar{c} q\bar{q}$, where $q=u,d$ quarks. We have discussed
 two competing dynamical frameworks for the $\psi(4230)$, based on the molecular and compact diquark-antidiquark  pictures. Given sufficient details of
 these frameworks, and the composition of this state in the underlying constituents, the decay amplitudes can, in principle,  be calculated.
 Instead, we have introduced a data-driven parametrization of the direct-production amplitude in terms of the leptonic coupling of the $\psi(4230)$, providing a practical way to quantify the short-distance overlap associated with the Higgs--charm interaction. Our calculations are carried out in the lowest order in $\alpha_s$ and hence are subject to higher order QCD corrections, such as the
 ones undertaken in the case of the $H \to J/\psi \gamma$ decay.  We obtain ${\cal B}_{\rm SM}( H \to \psi(4230)  \gamma)=4.88^{+4.71}_{-4.52}\times 10^{-7}$,
with the uncertainties dominated by the present experimental uncertainty on the leptonic decay width of $\psi(4230)$, which we expect will be greatly reduced  by
precision measurements at the BESIII and BelleII experiments. The indirect contribution is numerically dominant, while its interference with the direct amplitude provides sensitivity to the charm Yukawa coupling $y_c$. Possible deviations due to the multiplicative renormalization of the direct amplitude are briefly discussed and illustrated numerically.

\vspace*{3mm}
Exclusive radiative decays of the Higgs are difficult to measure due to their small branching ratios.  The currently allowed ranges for the Yukawa coupling $y_c$
obtained by the ATLAS and CMS experiments quantify this. The only possibility to measure them in the near future is at the High-Luminosity LHC (HL-LHC), with the 
HL-LHC-upgrade  currently in full force at CERN. In the more distant future, a high-energy version of the LHC is envisaged, called
the High-Energy LHC (HE-LHC). In a dedicated study prospecting the Higgs physics at these machines~\cite{Cepeda:2019klc}, the inclusive production of the
Higgs boson in $p + p \to H + X$ is estimated. The projected number of the Higgs bosons is reported as $N_{14}=1. 92 \times 10^8$ and
$N_{27}=2.5  \times 10^9$, where $N_{14}$ assumes an integrated luminosity of $3 {\rm ab}^{-1}$ at the center-of-mass proton-proton energy of 14 TeV,
and $N_{27}$ assumes an integrated luminosity of $15 {\rm ab}^{-1}$ at a center-of-mass proton-proton energy of 27 TeV. An updated estimate of
$N_{14}=1. 86 \times 10^8$  in the PDG Higgs review~\cite{ParticleDataGroup:2026mpi} is very similar. In the very distant future, the proposed Fcc-hh collider
is the ultimate Super-Higgs factory, with an anticipated yield of $2.8 \times 10^{10}$ Higgs bosons (assuming $\sqrt{s}= 100$ TeV and an integrated luminosity of
30 ab$^{-1}$)~\cite{FCC:2025lpp}. In all likelihood, searches of the rare Higgs decays will
require AI-techniques, developed to search for sparse events, early forms of which based on machine-learning  are already {\it en vogue} in the analysis of the current LHC data.

\vspace*{3mm}
  Final remark: In addition to the $\psi(4230)$ discussed here, there are other candidate exotic vector states present in the data in the mass range between 4150 MeV and 4660 MeV~\cite{ParticleDataGroup:2026mpi}.  They are anticipated in the diquark-antidiquark picture, as noted earlier. Exclusive radiative Higgs decays to these vector states can be worked out along the lines discussed here  for the decay $H \to \psi(4230) \gamma$, once they are established and their leptonic decay widths measured.

\section*{Acknowledgements}
\vspace*{-3mm}
One of us (AA) would like to thank the CERN-TH Division for the kind hospitality. We thank Alexander Parkhomenko for critically reading the manuscript and
helpful comments.


\end{document}